\documentclass[preprint,showpacs,preprintnumbers,amsmath,amssymb]{revtex4}

\usepackage{graphicx}
\usepackage{dcolumn}
\usepackage{bm}

\begin{document}

\title{Accurate discretization of advection-diffusion equations}
\author{R. Grima$^{1}$ and T. J. Newman$^{1,2}$}  

\affiliation{$^{1}$ Department of Physics and Astronomy, 
Arizona State University, Tempe, AZ 85284 \\
$^{2}$ School of Life Sciences, Arizona State University,
Tempe, AZ 85284}

\vspace{5mm}
\begin{abstract}
We present an exact mathematical transformation which converts a wide
class of advection-diffusion equations into a form allowing simple and
direct spatial discretization in all dimensions, and thus the
construction of accurate and more efficient numerical algorithms.
These discretized forms can also be viewed as master equations which
provides an alternative mesoscopic interpretation of
advection-diffusion processes in terms of diffusion with spatially
varying hopping rates.
\end{abstract}
\vspace{2mm} 
\pacs{02.70.Bf, 47.27.-i}

\maketitle

\section{Introduction}

Advection-diffusion equations (ADEs) describe a broad class of
processes in the natural sciences. As their name implies, they provide
a continuum (macroscopic) representation of systems whose underlying
dynamics combines Brownian motion (diffusion) with some form of
deterministic drift (advection).  In this paper we shall consider ADEs
of the general form
\begin{equation}
\label{ade-gena}
\partial _{t}\rho = \nabla \cdot D \nabla \rho - \nabla \cdot \rho {\bf v} \ . 
\end{equation} 
The field $\rho$ typically 
describes the number density of ``particles'' which, depending
on the application, can range from electrons in a plasma, to chemical
molecules advected in solution, to colloidal particles, to biological
cells moving along chemical gradients. In principle the diffusion
coefficient $D({\bf x},t)$ and the velocity field ${\bf v}({\bf x},t)$
can depend on the density field $\rho$.  An idea
of the ubiquity of ADEs can be gauged from their diverse applications
to traditional physics, soft matter systems, and biology.  A small
subset of examples are magnetic fusion plasmas \cite{arter}, cosmic
ray streaming \cite{skilling, ryu}, electrons in weakly ionized gases
\cite{carron}, microemulsions under shear flow \cite{gonnella},
chemical kinetics in driven systems \cite{schmidt, edwards},
hydrodynamics and chemotaxis of bacterial colonies \cite{lega, ks2},
phase field dynamics in directional solidification \cite{rousseau},
and a wide array of tracer diffusion problems (for example
\cite{tracer}).

It is generally not possible to analytically solve ADEs, especially
since they often appear within sets of non-linear coupled equations.
For this reason, great emphasis has been placed on numerical
integration methods, typically based on finite differences.  It has
been found that the advection term, despite its apparent simplicity,
is extremely troublesome to handle \cite{leveque1}. There are two major
challenges: {\it stability}, which can be improved using a range of
implicit methods, and {\it accuracy}, which is a delicate issue,
requiring the ``best possible'' form of spatial discretization. Regarding the
issue of stability, many schemes are in use, such as the
Crank-Nicholson/ADI and fractional step methods
\cite{leveque1, leveque2},
and the Lax-Wendroff method \cite{dehghan}. The issue of accuracy has received
somewhat less attention with two spatial discretization schemes (and their
immediate variants) commonly in use: these are the simple Taylor
expansion \cite{arter, othmer} and the ``upwind'' scheme 
\cite{garcia, painter}. One of the main
results of this paper is the derivation of a new discretization scheme
which is physically appealing, simple to apply in all dimensions, and
more accurate than those currently in use.

\section{A simple example}

To begin, let us present the key idea in the context of a simple ADE,
namely, a one-dimensional system with a velocity function proportional
to the spatial derivative of a scalar potential $\phi (x,t)$. Thus, we
consider
\begin{equation}
\label{ade1}
\partial _{t}\rho = D_{0}\partial _{x}^{2}\rho - \alpha \partial _{x}
(\rho \partial _{x}\phi) \ , 
\end{equation} 
where $D_{0}$ and $\alpha $ are constants.

Most numerical algorithms designed to integrate an equation such as
(\ref{ade1}) treat the diffusion and advection terms separately
\cite{arter, leveque1, painter}.  The difficulties arise in finding
a discretization for the latter term.  In doing so, two fundamental
properties of the equation must be exactly maintained. These are the
non-negativity of $\rho$ and its spatial conservation:
\begin{equation}
\label{cons}
\int dx \ \rho (x,t) = {\rm const} \ .
\end{equation} 

As an illustration, let us write down a common spatial discretization
using simple Taylor expansion, which is used both for explicit Euler
schemes \cite{edelstein}, and as the basis for more advanced implicit
algorithms \cite{garcia, leveque1, othmer}.  We replace the continuous
function $\rho (x,t)$ by a set of functions $\lbrace \rho _{i}(t) \rbrace $
defined on a regular grid with lattice spacing $h$. The equation of
motion for $\rho _{i}$ is written using centered spatial derivatives:
\begin{equation}
\label{ade1-disca}
\frac{d\rho _{i}}{dt} = \frac{D_{0}}{h^{2}}(\rho _{i+1}+\rho_{i-1}-2\rho_{i})
-\frac{\alpha}{4h^{2}}\left [\rho_{i+1}(\phi _{i+2}-\phi _{i}) 
-\rho_{i-1}(\phi _{i}-\phi_{i-2}) \right ] \ .
\end{equation} 
It is noteworthy that this simple scheme requires knowledge of the
scalar field $\phi$ at next-nearest neighbor grid points rather than
neighboring grid points. For future reference we rewrite this discrete
equation in the following manner:
\begin{equation}
\label{ade1-discb}
\frac{d\rho _{i}}{dt} = \frac{1}{h^{2}}\left \lbrace 
\rho _{i+1}\left [D_{0}-\frac{\alpha}{4}(\phi _{i+2}-\phi _{i})\right ]
+\rho _{i-1}\left [D_{0}+\frac{\alpha}{4}(\phi _{i}-\phi_{i-2})\right ]
-\rho _{i}[2D_{0}] \right \rbrace \ ,
\end{equation} 
which we shall hereafter refer to as the ``linear centered
discretization'' (LCD) (and which resembles the backward Euler scheme
used for simple advection problems \cite{leveque1}).

We now turn to a new discretization scheme which emerges from a simple
mathematical transformation of the ADE (\ref{ade1}). Defining $\gamma
= \alpha/2D_{0}$ it can be verified by direct differentiation that
Eq.(\ref{ade1}) may be written as
\begin{equation}
\label{ade1-exp}
\partial _{t}\rho = D_{0}\left [ 
e^{\gamma \phi} \ \partial _{x}^{2}(\rho e^{-\gamma \phi}) 
- e^{-\gamma \phi}\rho \ \partial _{x}^{2}(e^{\gamma \phi}) \right ] \ . 
\end{equation} 
A similar transformation involving exponential functions is known for
Fokker-Planck equations \cite{risken}.  The simple ADE given in
(\ref{ade1}) can indeed be formally interpreted as such an equation,
although the physical origin is quite different. We will shortly be
considering more general ADEs in which the diffusion coefficient and
velocity function can be functions of the density $\rho$. Clearly then
the simple correspondence with Fokker-Planck equations breaks down,
although we are still able to achieve a transformation of the kind
given above.  The crucial feature of Eq.(\ref{ade1-exp}) is that spatial
derivatives only enter in the form of a second derivative $\partial
_{x}^{2}$ which is straightforward to discretize. Using the simplest
such discretization we immediately have
\begin{equation}
\label{ade1-exp-disca}
\frac{d\rho _{i}}{dt} = \frac{D_{0}}{h^{2}}
\left [ \rho_{i+1}e^{-\gamma (\phi _{i+1}-\phi _{i})}
+\rho _{i-1}e^{-\gamma (\phi _{i-1}-\phi _{i})}
-\rho _{i}(e^{-\gamma (\phi _{i}-\phi _{i+1})}+
e^{-\gamma (\phi _{i}-\phi_{i-1})})
\right ] \ .
\end{equation} 

There are a number of points to make concerning this equation.  First,
in contrast to the LCD (\ref{ade1-disca}), the
scalar field appears in a non-linear fashion, and is
sampled at nearest-neighbor positions. Second, the new equation is of
the same form as a master equation \cite{othmer, risken, gardiner}.
Within this analogy one can think of $\rho _{i}$ as the probability that
a fictitious particle is located at grid position $i$. The transition rate for
the particle to hop from grid point $i$ to a neighboring point $j$ is
of the Arrhenius form
\begin{equation}
\label{transition}
W_{i \rightarrow j}=(D_{0}/h^{2})\exp[-\gamma (\phi _{i}-\phi _{j}) ] \ .
\end{equation} 
Given this formal analogy with a master equation for a probability
function, one immediately sees that Eq.(\ref{ade1-exp-disca}) exactly
maintains conservation of the function $\rho$ (normalization of
probability) and its non-negativity. Due to this analogy we hereafter
refer to Eq.(\ref{ade1-exp-disca}) as the ``master equation
discretization'' (MED).

Our numerical work (see section V) shows that the MED is more accurate
than the LCD and other popular discretizations. To appreciate the
underlying reason for this, it is helpful to consider the case of
$\gamma \delta \phi \ll 1$ in which case we can expand the exponential
functions in Eq.(\ref{ade1-exp-disca}) to first order. One then finds
\begin{eqnarray}
\label{ade1-exp-discb}
\nonumber
\frac{d\rho _{i}}{dt} = \frac{1}{h^{2}}\Bigl \lbrace 
\rho _{i+1}\left [D_{0}-\frac{\alpha}{2}(\phi _{i+1}-\phi _{i})\right ]
& + & \rho _{i-1}\left [D_{0}+
\frac{\alpha}{2}(\phi _{i}-\phi_{i-1})\right ] \\
& - & \rho_{i}\left [ 2D_{0} - \frac{\alpha}{2}
(2\phi _{i} - \phi_{i+1} - \phi_{i-1}) \right ] \Bigr \rbrace \ .
\end{eqnarray} 
Comparison of this form with Eq.(\ref{ade1-discb}) gives useful
insight into the potential weakness of the LCD. Namely, it neglects an
important curvature term in the scalar field. In fact, this omission
is directly related to artificial (or ``numerical'') diffusion, which is a
common failing of other discretization schemes, most notably, the
``upwind'' scheme \cite{garcia, leveque1, painter}. The linear scheme
given above in Eq.(\ref{ade1-exp-discb}) can of course be regarded as
one of many possible linear discretizations, but without the
derivation given here one would have no {\it a priori} reason to
prefer it over forms such as the LCD, since they both have
non-vanishing second-order errors in space. Continuing the expansion of the
exponential terms in powers of $\alpha $ yields crucial non-linear
corrections to
Eq.(\ref{ade1-exp-discb}) which have no analogy within linear
discretization schemes. As shall be seen below, the MED is easily
formulated for the $d$-dimensional extension of Eq.(\ref{ade1}) as
well as for a range of more general ADEs.

\section{The general case}

Consider the general ADE in $d$-dimensions given in
Eq.(\ref{ade-gena}).  We shall now proceed to transform this equation
into a form amenable to the MED.  In one dimension we shall find that
this is possible for general functions $D$ and $v$. In higher dimensions
the vectorial nature of the velocity field will place a constraint on
the transformation.

Let us introduce two scalar functions $f({\bf x},t)$ and $g({\bf
x},t)$ defined via the relations
\begin{eqnarray}
\label{dfg}
D & = & f g \ , \\
\label{vfg}
{\bf v} & = & g \nabla f - f \nabla g \ .
\end{eqnarray} 
Then the ADE (\ref{ade-gena}) has the explicit form
\begin{equation}
\label{ade-genb}
\partial _{t}\rho = \nabla \cdot [fg \nabla \rho] - \nabla \cdot [\rho 
(g \nabla f - f \nabla g)] \ . 
\end{equation}
By direct differentiation one can show that this equation may be
rewritten as
\begin{equation}
\label{ade-genc}
\partial _{t}\rho = f\nabla ^{2}(g\rho) - g\rho\nabla ^{2}f \ .
\end{equation}
Once again, we see that the spatial derivatives appear only as
Laplacians, which allows us to immediately write down a simple
discrete form. Let us define the discrete Laplacian via
\begin{equation}
\label{disclap}
\nabla ^{2}Q({\bf x}) = \frac{1}{h^{2}}{\sum \limits _{j}}'(Q_{j}-Q_{i}) \ ,
\end{equation}
where the sum is over nearest neighbors $j$ of the grid point $i$,
which corresponds to the continuum position ${\bf x}$. Then the MED
corresponding to Eq.(\ref{ade-genc}) is
\begin{equation}
\label{ade-gen-disc}
\partial _{t}\rho _{i} = {\sum \limits _{j}}' [W_{j \rightarrow i} \ \rho_{j}
- W_{i \rightarrow j} \ \rho_{i} ] \ ,
\end{equation}
where the transition rate for ``hopping'' from site $i$ to site $j$ is
\begin{equation}
\label{transition-gen}
W_{i \rightarrow j} = f_{j} \ g_{i}/h^{2} \ .
\end{equation}
Having formulated the MED in this general manner, let us examine some
particular cases. We stress that once the functions $f$ and $g$ are
determined the discrete algorithm is completely defined via the
transition rate given above.

First, we consider one dimension. In this case it is possible to
integrate Eqs. (\ref{dfg}) and (\ref{vfg}) exactly to find the
necessary auxiliary functions $f$ and $g$ in terms of the physically
relevant diffusion coefficient and velocity. One finds
\begin{equation}
\label{fandg}
f(x,t)=C\sqrt{D(x,t)}\exp(S), \ \ \ S(x,t)=\frac{1}{2}\int \limits ^{x}
dx' \ \frac{v(x',t)}{D(x',t')} \ ,
\end{equation}
with $g$ then given trivially from (\ref{dfg}). 
The transition rate is easily evaluated from (\ref{transition-gen}) 
to give
\begin{equation}
\label{transition-gen-1d}
W_{i \rightarrow j} = \frac {\sqrt{D_{i}D_{j}}}{h^{2}} \ 
\exp[-(S_{i}-S_{j})] \ .
\end{equation}
A non-trivial application of this general solution would be
advection-diffusion in the kinetic theory of gases where 
the diffusion coefficient is non-constant, and actually depends on the 
density as $D \propto 1/\rho $ \cite{inverserho}. 
In higher dimensions a general solution for $f$ and $g$
is not possible. Solvable cases will rely on special conditions
for $D$ and ${\bf v}$ reminiscent of the potential conditions for the existence
of steady-state solutions to the multi-variate Fokker-Planck equation
\cite{risken, gardiner}.

For many problems the diffusion coefficient is constant ($D_{0}$) and
the velocity function is associated with a scalar potential via ${\bf
v}=\alpha \nabla \phi$. In these cases, the analysis leading to
Eq.(\ref{transition}) is easily generalized to $d$ dimensions and one
finds the discrete equation (\ref{ade-gen-disc}) with
\begin{equation}
\label{transition-gend}
W_{i \rightarrow j}=(D_{0}/h^{2})\exp[-\gamma (\phi _{i}-\phi _{j}) ] \ ,
\end{equation}
where we remind the reader that $\gamma = \alpha /2D_{0}$.  As found
in one dimension, this scheme includes important curvature terms, even
within a linear approximation, which are absent in conventional LCD
algorithms. Numerical analysis shows such terms
to be essential in regions where $\phi $ has maxima or minima.

The MED scheme encapsulated in Eqs. (\ref{ade-gen-disc}) and
(\ref{transition-gend}) can be used to model more complicated ADEs in
which there is non-linear feedback.  An interesting example of this is
the continuum theory of group dynamics, in which a non-linear and
non-local feedback mechanism is imposed via the velocity potential
\cite{levin, flierl}. In particular one has
\begin{equation}
\label{group}
\phi (x,t) = \int d^{d}x' V({\bf x}-{\bf x'})\rho ({\bf x'},t) \ ,
\end{equation}
where $V$ is analogous to
a potential, and is responsible for long-range attraction and short-range
repulsion of individuals.
If $V$ is a Dirac $\delta$-function then $\phi \propto \rho $. Such
models are used to describe density-dependent dispersal in population
dynamics \cite{levin} and have recently been shown to arise from
excluded volume effects in models of interacting cellular systems
\cite{ng}.  A second well-known example is the Keller-Segel model for
chemotactic motion \cite{ks2}.  Here, the potential $\phi$ represents
the chemoattractant concentration field and is coupled to the cell
density field $\rho$ via
\begin{equation}
\label{ks}
\partial _{t}\phi = \nu \nabla ^{2}\phi - \lambda \phi + \beta \rho \ ,
\end{equation}
where $\nu $, $\lambda $ and $\beta $ are the diffusion constant for
the chemical field and its rate of degradation and production
respectively.  This equation is easily discretized and the resulting
discrete chemical concentration field may be inserted into the
transition rate (\ref{transition-gend}) allowing a straightforward
scheme for integration of the cell density.

\section{Fine-tuning the MED algorithm}

From numerical investigations (see next section) we have found that the
MED is generally far more accurate than both the LCD and upwind schemes.
In regions where the velocity function has strong spatial variation, 
the MED does an excellent job in predicting the correct
density even for grid scales approaching the scale of
variation of the velocity. However, in the ``simpler case'' when dynamics
are dominated by advection in a region of quasi-constant velocity, the
MED fares less well. This problem can be traced back to the exponential
weights yielding, in regions of constant velocity, an over-estimated
drift velocity. In terms of a hopping process, the bias in hopping rates
between neighboring sites is proportional to ${\rm sinh} (\gamma \delta \phi)$,
whereas the correct drift velocity is simply proportional to 
$\gamma \delta \phi$. 

We discuss here two straightforward extensions to MED which alleviate 
this problem, but also lead to slightly less accurate algorithms in the
``non-trivial'' regions where the velocity is strongly varying.
Both extensions amount to a renormalization of the hopping rates. An
ideal algorithm would be a hybrid, using the original MED and either of the
following extensions in appropriate regions. We will not discuss such hybrid
schemes here since their form will be strongly dependent on 
actual applications.

For simplicity let us consider again the one-dimensional ADE given in
Eq.(\ref{ade1}). The MED scheme for this case in given in 
Eq.(\ref{ade1-exp-disca}), where the transition rate from site $i$ to 
neighboring site $j$ has the explicit form
\begin{equation}
\label{transitionrep}
W_{i \rightarrow j}=(D_{0}/h^{2})\exp[-\gamma (\phi _{i}-\phi _{j}) ] \ .
\end{equation} 
It is clear from (\ref{transitionrep}) 
that the effective drift velocity arising from the bias in 
hopping rates between $i$ and $j$ is
\begin{equation}
\label{bias}
v_{\rm eff} = h (W_{i \rightarrow j}-W_{j \rightarrow i}) =
(2D_{0}/h) {\rm sinh} [\alpha (\phi _{j}-\phi _{i})/2D_{0}] \ ,
\end{equation}
where we have reinstated $\gamma = \alpha/2D_{0}$ for clarity.
The correct drift velocity between these two points is simply
$\alpha (\phi _{j}-\phi _{i})/h$ which is recovered if the grid scale is
small (or else the velocity potential is slowly varying). 

In order to correct the MED algorithm one may either renormalize the effective
diffusion coefficient (which is the pre-factor of the exponential weight)
or else renormalize the parameter $\gamma $ which appears in the argument
of the exponential. In the former case one has, on fitting the drift
velocity to its correct value, the effective diffusion coefficient
\begin{equation}
\label{deff}
D_{\rm eff} = D_{0}\frac {\alpha \delta \phi/2D_{0}}
{{\rm sinh}(\alpha \delta \phi/2D_{0})} 
\end{equation}
which leads to the MED transition weight taking the ``Fermi-Dirac'' (FD) form
\begin{equation}
\label{fermi-dirac}
W_{i \rightarrow j}=\left (\frac {D_{0}}{h^{2}} \right )
\frac {\alpha(\phi_{i}-\phi_{j})/D_{0}}
{\exp[\alpha (\phi _{i}-\phi _{j})/D_{0}]-1} \ .
\end{equation} 

The alternative is to correct the drift velocity by adjusting $\gamma $,
which leads to 
\begin{equation}
\label{gammaeff}
\gamma_{\rm eff} = \frac {1}{\delta \phi}{\rm sinh}^{-1}\left (
\frac {\alpha \delta \phi}{2D_{0}} \right ) \ . 
\end{equation}
Writing the inverse hyperbolic function in terms of a 
logarithm leads to the MED transition rate taking the ``square root'' (SR) 
form
\begin{equation}
\label{sqrtform}
W_{i \rightarrow j}=(D_{0}/h^{2})\left \lbrace \left [1+ \left ( \frac {\alpha 
(\phi _{i}-\phi _{j})}{2D_{0}} \right )^{2} \right ] ^{1/2} - 
\left ( \frac {\alpha (\phi _{i}-\phi _{j})}{2D_{0}} \right ) \right 
\rbrace \ .
\end{equation} 

Numerically one finds that the FD form (\ref{fermi-dirac}) 
is generally more accurate than the SR form (\ref{sqrtform}),
and that both are superior to the LCD and upwind schemes. As already 
mentioned, the original MED scheme defined by Eq.(\ref{transitionrep})
is the best of all the schemes described when the velocity field is
strongly varying, and/or during asymptotic relaxation of the
density field to its steady-state.

\section{Numerical work}

We have made a careful numerical analysis of the simple
one-dimensional ADE given in Eq.(\ref{ade1}), along with its
two-dimensional extension. Since we wish to gauge the accuracy of our
new scheme, we have compared the MED scheme (\ref{ade1-exp-disca}), 
and its variants (the MED(FD) given in (\ref{fermi-dirac}), 
the MED(SR) given in (\ref{sqrtform}), and the linearized MED, 
denoted by MED(LIN), given in (\ref{ade1-exp-discb})), 
with both the LCD and upwind schemes \cite{leveque1}. In one dimension 
we use a static velocity potential given by
$\phi(x)=[1+{\rm cos}(2\pi nx/L)]/2$ with $n=16$ and $L=12.8$. The initial
density function is taken to be uniform in the region $x \in (-3,3)$ and
zero otherwise. The density is normalized to unity and periodic boundary
conditions are enforced. This set-up provides
a challenging test of all the schemes since the velocity field is a
strongly varying function of position. Furthermore, we challenge the methods
by using the parameter values $D_{0}=1.0$ and $\alpha = 5.0$ (Figure 1) and
$\alpha = 20.0$ (Figure 2), which correspond to moderate to high grid 
Peclet numbers \cite{arter} at the grid scales of interest. Here, the 
largest Peclet numbers are approximately given by $2\alpha h$ and so
vary between 0.25 and 8 for the data shown in Figures 1 and 2. The dynamics
consists of a rapid transient phase where the density field adapts to the
periodic structure of the velocity field, followed by a slower relaxation
towards the steady-state. Thus, the numerical analysis probes each scheme's
ability to track rapid advective motion and diffusive relaxation around
maxima and minima of the velocity field.

In order to assess the accuracy of the methods we first run all schemes
at a very small grid size of $h=0.00625$, using an explicit temporal
scheme with $\delta t=10^{-6}$. 
Very good agreement is found among all the schemes
and the solution is denoted ``exact.'' We then run all the schemes at
larger grid scales using $\delta t = 10^{-4}$, 
and dynamically compare the approximate solutions
with the exact one. This is gauged using the relative error, which is
defined via
\begin{equation}
\label{relerror}
E(t)=\frac {\sum \limits _{i} [\rho _{i}(t) - \rho_{i, \rm {exact}} (t)]^{2}}
{\sum \limits _{i} \rho_{i, \rm {exact}} (t)^{2}}
\end{equation} 
Note, that $\delta t$ is chosen small enough such that any differences between
our first-order temporal discretization for LCD and second-order 
schemes (in the temporal dimension) such as Crank-Nicholson or Lax-Wendroff
are negligible.
Figures 1(a)-(e) show $E(t)$ for grid scales $h=0.025, \ 0.05, \ 0.1, \ 0.2$,
and $0.4$ respectively, for $\alpha=5.0$. The entire dynamical evolution 
up to the steady state is shown. In the first four panels we clearly see
that the MED and its (nonlinear) 
variants give a relative error approximately 10 times
less than the LCD and UW schemes. (Note UW does not appear in 1(a) since
its error is too large to be usefully included in the figure.) The relative
errors of all the schemes increases roughly 
by a factor of 10 as the grid scale
is doubled. Panel 1(e) shows the breakdown of all the schemes at the
scale $h=0.4$ which is comparable to the period of the velocity field.
By ``breakdown'' we mean a relative error of 10$\%$ or more.
To give an idea of the spatial form of the density field near the steady-state
we show in Figure 1(f) the exact density profile in a peripheral region,
along with the LCD and MED (FD) at a grid scale of $h=0.2$ 
for comparison. Note the LCD fails to capture the magnitude of the maximum
density, and also becomes negative at some grid points.

In a similar fashion, figures 2(a)-(d) show $E(t)$
for $h=0.025, \ 0.05, \ 0.1$, and $0.2$ respectively, for $\alpha=20.0$.
As before the non-linear 
MED schemes perform far better than the LCD and UW, meaning
the relative error is roughly 10 times smaller for a given grid scale.
Note also that the MED(FD) and MED(SR) algorithms perform better than MED
during the transient period, as expected.
All schemes break down for $h=0.2$. In Figure 2(e) we show the
exact density profile close to the steady-state, compared with the 
MED and LCD schemes for $h=0.1$. Again, the LCD shows negative values and
fails in the vicinity of the density peaks. Figure 2(f) is the same except
the UW scheme is compared to the MED. The UW scheme is designed to give
non-negative densities, but has high (artificial) ``numerical diffusion''
which inflate the width of the density peaks. 

We have performed an exactly analogous numerical examination in two
dimensions. We integrated the 2D generalization of Eq.(\ref{ade1}) using
the potential $\phi(x,y)=[1+{\rm cos}(2\pi nx/L)][1+{\rm cos}(2 \pi ny/L)]/4$
with $n=16$ and $L=12.8$. We take $D_{0}=1.0$ and $\alpha = 10.0$. The
initial density function is uniform in a disk of radius 3.0 and zero
otherwise, and again normalized to unity. The ``exact'' density profile
is evaluated using $h=0.0125$ and 
$\delta t=0.25 \times 10^{-4}$. The two-dimensional
extensions of all six schemes are integrated for grid scales of 
$h=0.025, \ 0.05, \ 0.1$, and $0.2$ using $\delta t = 10^{-4}$. 
The relative error $E(t)$ for these
cases is shown in Figure 3(a)-(d), for a time period encompassing the
initial rapid adaptation to the potential followed by the early stages of
relaxation to the steady-state. As with one dimension, the MED and 
its (non-linear) variants
perform far better than the LCD and UW, with the pure MED scheme performing
best at later times. All schemes break down for $h=0.2$.
Direct comparison of the exact density profile, MED, 
and LCD is given in Figures 3(e) and (f), for $h=0.05$ and $h=0.1$ 
respectively,
along a one-dimensional cut ($y=0$) in a peripheral region of the density.
The MED shows excellent agreement, especially in the vicinity of the
density peaks. The LCD fails in the vicinity of the density peaks as expected. 

From this and similar numerical work we have concluded that the MED 
and its (non-linear) extensions are
superior spatial discretization schemes compared to the LCD and upwind
schemes. The MED works especially well in regions of large variation in the
velocity potential. Generally speaking, for a given error tolerance, the
MED and variants allow one to use grid scales at least two times larger
than traditional schemes, which translates into a saving of {\it at least}
a factor of 4 and 8 in computational cost for two and three dimensional
numerical analyses.

\section{Discussion and conclusions}

We end with some remarks on the non-linear transition rates of the
MED. In most applications the ADEs represent processes for which there
is no underlying lattice (e.g. cosmic ray diffusion \cite{ryu} 
or chemotactically moving cells \cite{ks2}).  When one discretizes the
continuum ADE one must therefore not regard the lattice version as
``more fundamental'' or ``more microscopic.'' It is simply a mathematical
analog of the original equation and identical in the limit of the
lattice spacing being taken to zero. This is a different situation
to that found for many models arising from solid state physics in
which there is an underlying crystal lattice, and for which the
discrete equation can often be regarded as more fundamental (or, at least,
more microscopic) than continuum models. Although the hopping process
encapsulated by the MED cannot be viewed as the
underlying microscopic dynamics, it is interesting that ADEs can be
accurately modeled by a process in which diffusion and advection are
non-linearly combined in Arrhenius transition rates. Figure 4
summarizes our understanding of
the algorithmic connections between ADE and the MED
discretization, in which a given ADE typically arises from a mean-field
approximation of a microscopic stochastic process which is not
constrained by a lattice.

Pragmatically one wishes to impose a ``large'' lattice scale
for numerical efficiency, while avoiding
the loss of accuracy. Algorithms which remain accurate for
larger lattice scales yield great computational speed-up in higher
dimensions, since the number of required grid points (and hence
computer operations) scales as $h^{-d}$.  We find that our new scheme
typically allows grid scales between 2-4 times larger than
traditional schemes, which in three dimensions allows a potential speed-up in
computation of one or two orders of magnitude. Naturally, our improved
spatial discretizations can be used in more advanced algorithms
which use implicit temporal methods and/or adaptive spatial grids.

In conclusion we have shown that a wide class of advection-diffusion
equations can be exactly rewritten in a form which immediately allows
a direct and simple spatial discretization in all dimensions.  
Our new discrete forms
contain important non-linear terms, which when linearized are seen to
be related to the curvature of the velocity potential, such terms
being absent in commonly used discretization schemes. We have shown
explicitly that these curvature effects are essential for accurate
integration of ADEs, both in one and two dimensions, and allow simple
algorithms to be constructed which are accurate for grid
scales up to the size of spatial variation in the velocity field. We
estimate that our new algorithm may allow a speed-up of ADE
computation by factors of 10 or more in three dimensions due to the
increased grid scale one can impose. The fact that ADE can be recast
as master equations also yields interesting physical insight into
their dynamics - namely that at mesoscopic scales the processes of
diffusion and advection may be modeled as a
non-linear combination within Arrhenius-like transition rates.

The authors gratefully acknowledge partial support from NSF award
DEB-0328267.

\bibliography{adv-diff}

\newpage

\noindent
{\bf Figure Captions}

\vspace{0.2in}

\noindent
\underline{Figure 1}: Data from numerical integration of Eq.(\ref{ade1}) 
using various schemes 
in one dimension, with $D_{0}=1.0$ and $\alpha = 5.0$. The particular form
of the velocity potential and the initial density profile are described
in section V. 
The time step is $\delta t = 10^{-4}$. Figures 1(a), (b), (c), (d),
and (e) show the relative error (\ref{relerror}) 
as a function of time for grid scales
of $h=0.025, \ 0.05, \ 0.1, \ 0.2$, and $0.4$ respectively. The methods used
are upwind (UW), LCD (\ref{ade1-disca}), linearized MED
(\ref{ade1-exp-discb}), MED (\ref{ade1-exp-disca}), ``Fermi-Dirac''
version of MED (\ref{fermi-dirac}), and ``square-root'' version
of MED (\ref{sqrtform}). Figure 1(f) compares the exact density profile
in the peripheral region $x \in (2,3.6)$ with both the MED(FD) scheme and
the LCD scheme at time $t=0.1$ using $h=0.2$. In Figures 1-3, time is
measured in units of $\delta t$, space in units of $h$, and the density
in dimensionless units.

\vspace{0.2in}

\noindent
\underline{Figure 2}: 
Same as Figure 1, but with $\alpha=20.0$.  Figures 2(a), (b), (c),
and (d) show the relative error (\ref{relerror}) 
as a function of time for grid scales
of $h=0.025, \ 0.05, \ 0.1$, and $0.2$ respectively.
Figure 2(e) compares the exact density profile
in the peripheral region $x \in (2,3.6)$ with both the MED scheme and
the LCD scheme at time $t=0.02$ using $h=0.1$. 
Figure 2(f) is the same as 2(e) but
compares the exact profile with both MED and UW.

\vspace{0.2in}

\noindent
\underline{Figure 3}: Data from 
numerical integration of the two-dimensional generalization
of Eq.(\ref{ade1}) using various schemes, 
with $D_{0}=1.0$ and $\alpha = 10.0$. The particular form
of the velocity potential and the initial density profile are described
in section V. 
The time step is $\delta t = 10^{-4}$. Figures 1(a), (b), (c),
and (d) show the relative error (\ref{relerror}) 
as a function of time for grid scales
of $h=0.025, \ 0.05, \ 0.1$, and $0.2$ respectively. The methods used
are two dimensional generalizations of 
upwind (UW), LCD (\ref{ade1-disca}), linearized MED
(\ref{ade1-exp-discb}), MED (\ref{ade1-exp-disca}), ``Fermi-Dirac''
version of MED (\ref{fermi-dirac}), and ``square-root'' version
of MED (\ref{sqrtform}). Figure 3(e) compares the exact density profile
along a cut ($y=0$) in the peripheral region $x \in (2,3.6)$ 
with both the MED scheme and the LCD scheme at time $t=0.01$ using
$h=0.05$. Figure 3(f) is the same as 3(e) except that a larger grid
scale of $h=0.1$ is used.
 
\vspace{0.2in}
\underline{Figure 4}: 
A schematic diagram summarizing the relationships between various
descriptions of advection-diffusion processes. The MED is a useful
mesoscopic description in terms of Arrhenius hopping rates, rather
than a reflection of the underlying dynamics. 

\newpage

\begin{figure}
\includegraphics[width=\linewidth]{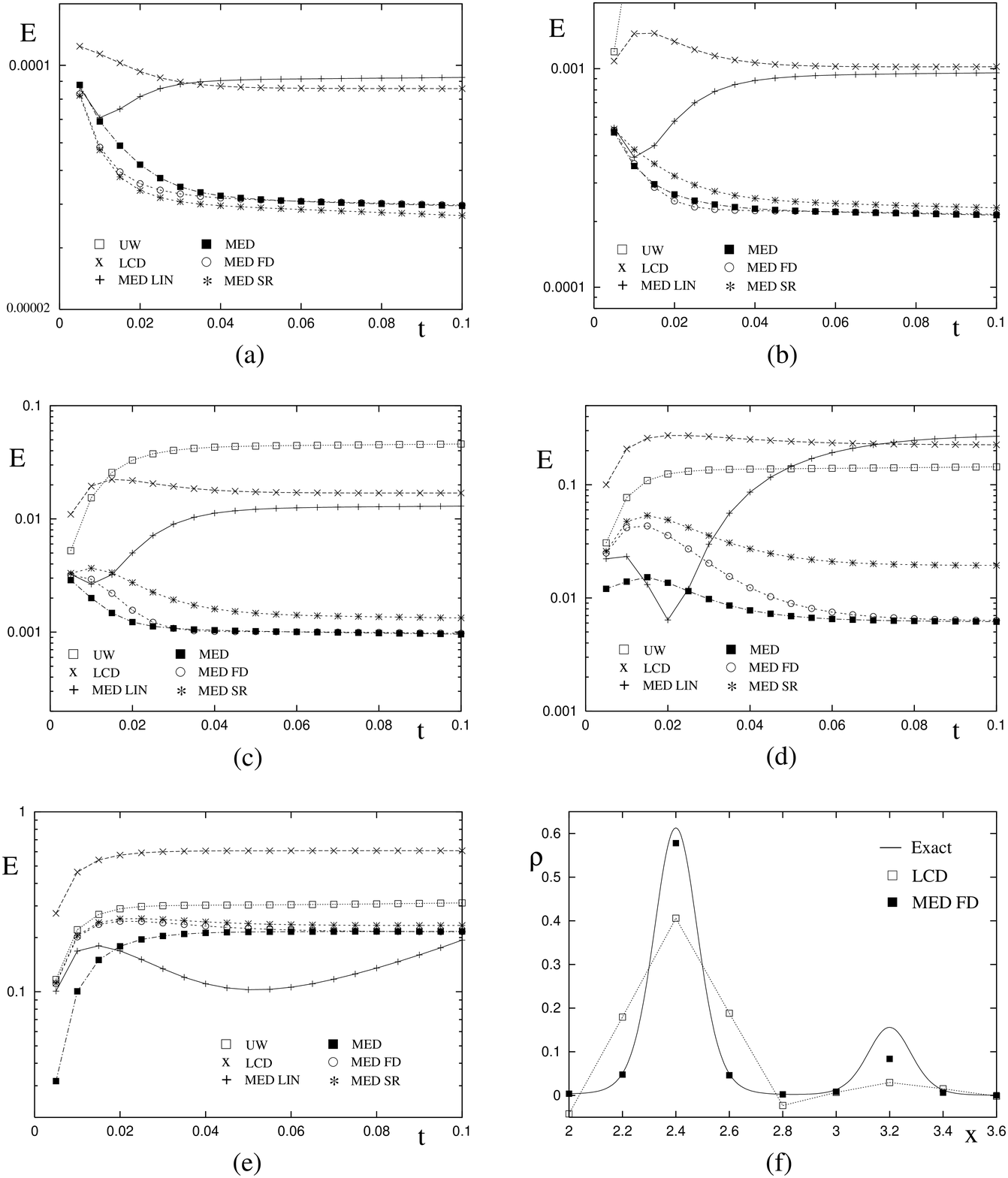}
\caption{}
\end{figure}

\begin{figure}
\includegraphics[width=\linewidth]{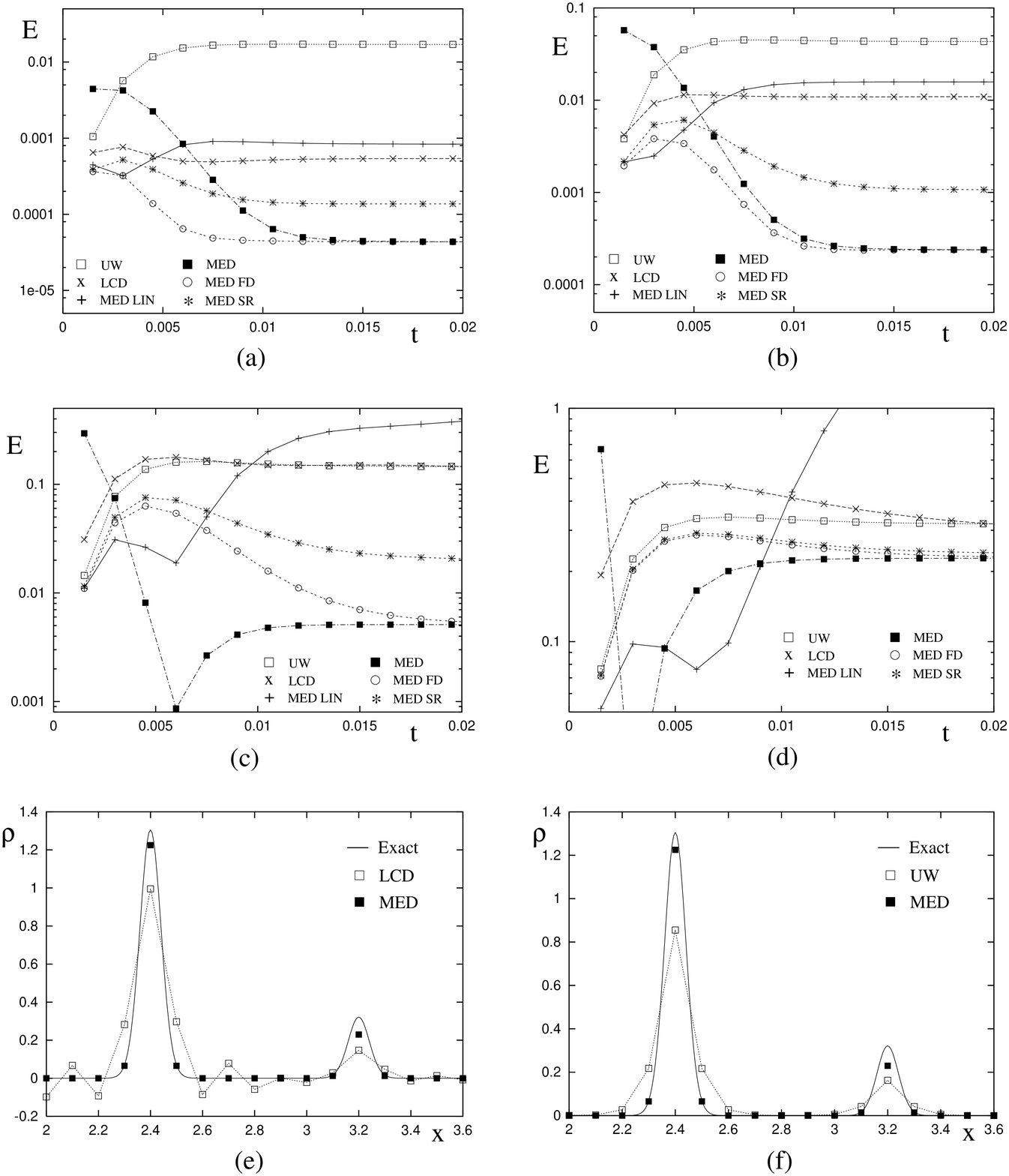}
\caption{}
\end{figure}

\begin{figure}
\includegraphics[width=\linewidth]{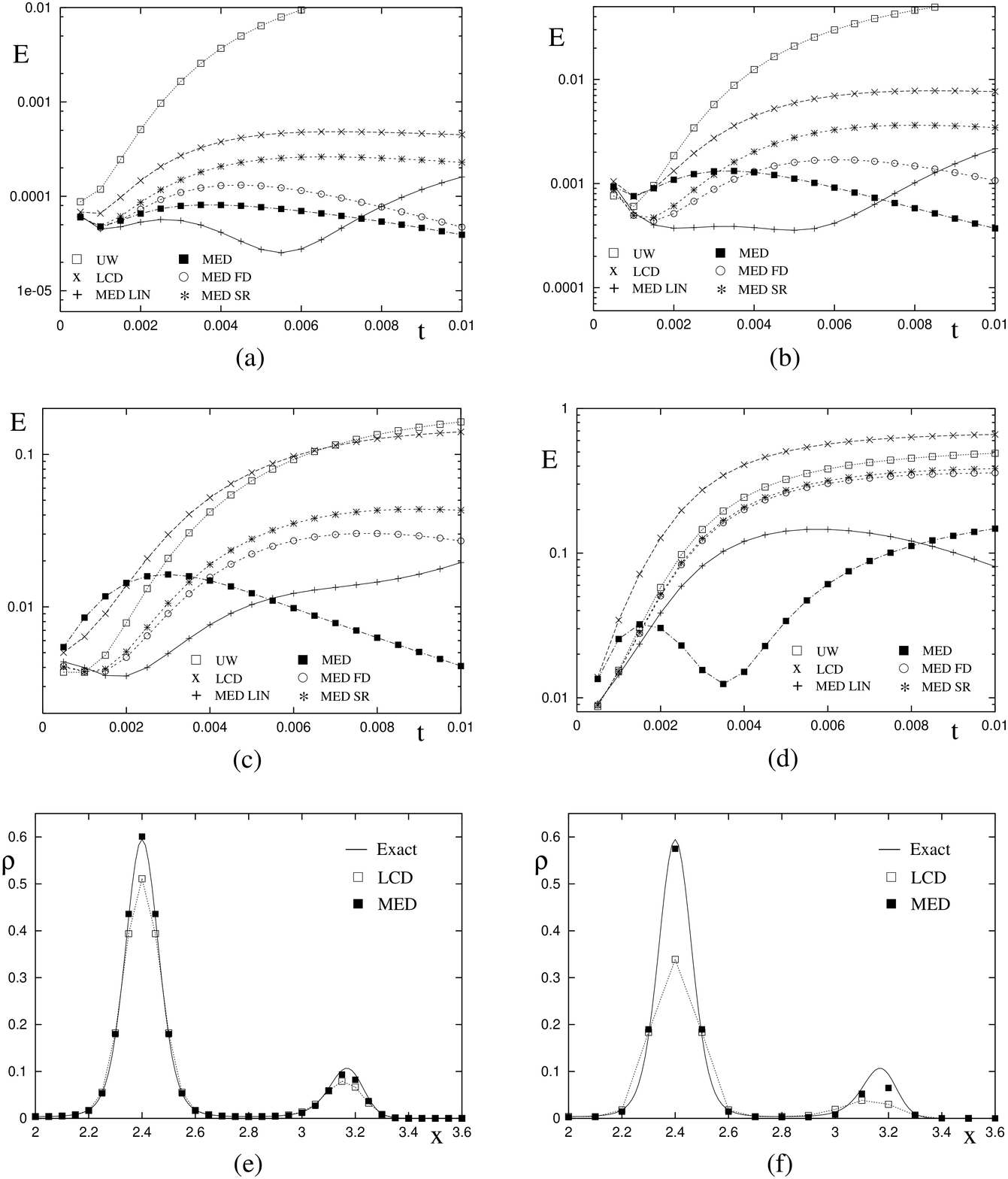}
\caption{}
\end{figure}

\begin{figure}
\includegraphics[width=6in]{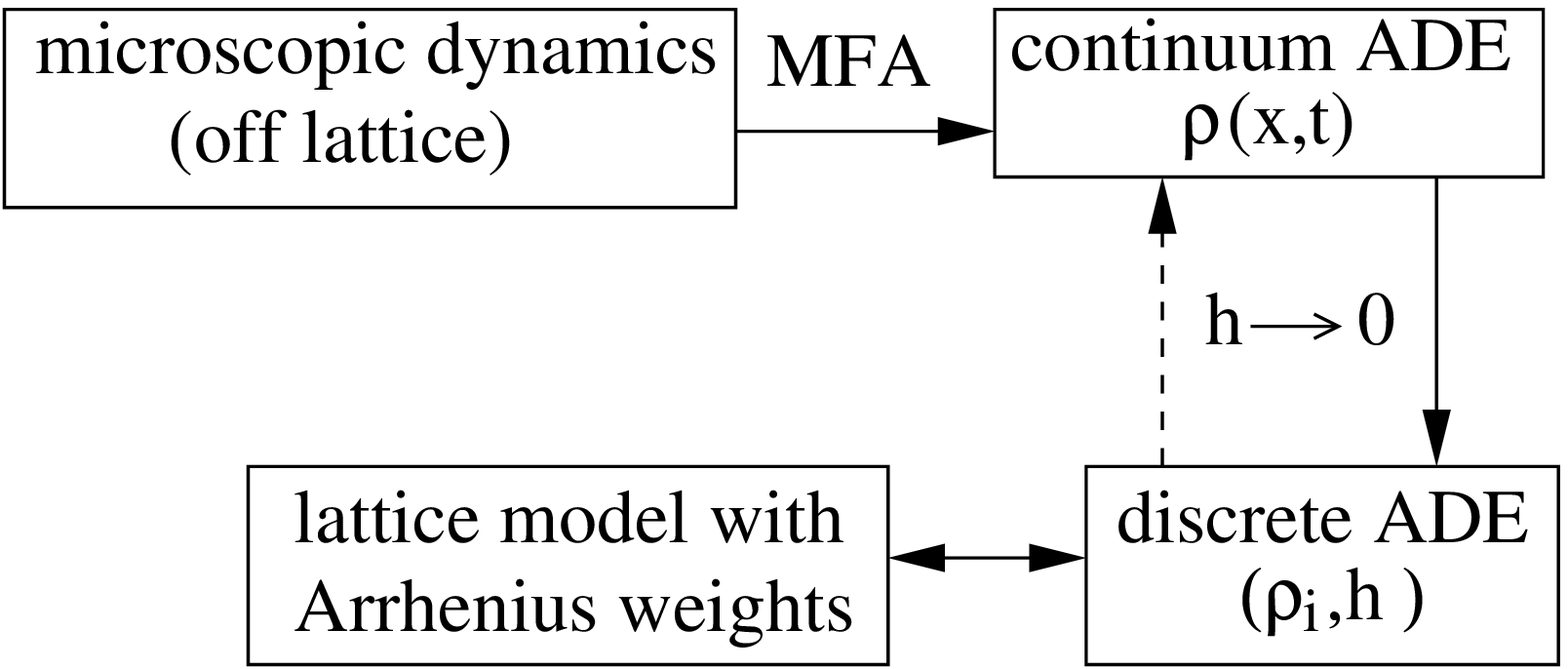}
\caption{}
\end{figure}

\end{document}